\documentclass{article}

\usepackage{amssymb,amsfonts,amsmath,stmaryrd}
\usepackage{cite,enumerate,float,indentfirst}
\usepackage{color}

\def\be{\begin{eqnarray}}
\def\ee{\end{eqnarray}}
\def\nn{\nonumber}

\def\p{\partial}

\definecolor{red}{rgb}{1,0,0}
\definecolor{orange}{rgb}{1,0.5,0}
\definecolor{violet}{rgb}{0.7,0,1}

\hoffset= - 1.0in         

\begin{document}

\hfill MIPT/TH- 08/19

\hfill ITEP/TH- 15/19

\hfill IITP/TH- 09/19

\bigskip

\centerline{\Large{ Pentad and triangular structures behind the Racah matrices
}}

\bigskip

\centerline{{\bf A.Morozov}}

\bigskip


{\footnotesize
\centerline{\it Moscow Institute for Physics and Technology, Dolgoprudny, Russia}

\centerline{\it ITEP, Moscow, Russia}

\centerline{\it    Institute for Information Transmission Problems, Moscow, Russia}
}

\bigskip

\centerline{ABSTRACT}

\bigskip

{\footnotesize
Somewhat unexpectedly,
the study of the family of twisted knots
revealed a hidden structure behind exclusive
Racah matrices $\bar S$, which control non-associativity
of the representation product in a peculiar channel
$R\otimes \bar R \otimes R  \longrightarrow R$.
These $\bar S$ are simultaneously symmetric and orthogonal,
and therefore admit two decompositions:
as quadratic forms, $\bar S \sim {\cal E}^{tr}{\cal E}$,
and as operators: $\bar T\bar S\bar T =  S T^{-1} S^{-1}$.
Here $\bar T$ and $T$ consist of the eigenvalues of the
quantum ${\cal R}$-matrices
in channels $R\otimes \bar R$ and $R\otimes R$ respectively,
$S$ is the second exclusive Racah matrix for
$\bar R\otimes R\otimes R \longrightarrow R$
(still orthogonal, but no longer symmetric),
and ${\cal E}$ is a {\it triangular} matrix.
It can be further used to construct the KNTZ evolution
matrix ${\cal B}={\cal E}\bar T^2{\cal E}^{-1}$,
which is also triangular and explicitly expressible through
the skew Schur and Macdonald functions -- what makes
Racah matrices calculable.
Moreover, ${\cal B}$ is somewhat similar to Ruijsenaars
Hamiltonian, which is used to define Macdonald functions,
and gets triangular in the Schur basis.
Discovery of this pentad structure
$(\bar T,\bar S,S,{\cal E},{\cal B})$, associated with
the universal ${\cal R}$-matrix,
can lead to further insights about representation theory,
knot invariants and Macdonald-Kerov functions.

}

\bigskip

\bigskip

Triangular structures are long believed to be intimately related to integrability.
For example, triangular are the pseudo-differential "dressing operators"
in the description of KdV/KP hierarchies \cite{dressing}.
In a complementary development,
triangularity in the form of Gauss decomposition proved to be crucial
for the free-field description of generic conformal theories \cite{GMMOS}.
Since then triangular transforms appeared in many places,
in particular, in the description of Macdonald-Kerov functions,
of their "generalized" multi-time deformations and, somewhat surprisingly,
in the study of "evolution" along the family of twist knots --
which had direct implication to the theory of Racah matrices ($6j$-symbols) \cite{Racah},
and allowed to explicitly calculate some of them, what remained an unsolvable problem
for quite a long period of time.
A purpose of this short note is a brief review of this newly emerging
and promisingly diverse field.

\bigskip

{\bf 1.} Another example of the fundamental importance is description of Macdonald \cite{Macd}
and, more generally, Kerov functions \cite{Kerov,Kerov1}
as triangular transformation of Schur functions,
see \cite{MMKerov} for a recent reminder.
In this case one just applies a triangular orthogonalization procedure
for the scalar product
\be
\Big<p^{\Delta }|p^{\Delta'}\Big>^{(g)} = z_\Delta \cdot \delta_{\Delta,\Delta'} \cdot
\left(\prod_{i=1}^{l_\Delta} g_{\delta_i}\right)
\ee
on the space of time-variables $p_k$.
Here $p^\Delta$ is the monomial basis,
$p^\Delta = \prod_{i=1}^{l_\Delta} p_{\delta_i}$, labeled  by Young diagrams
$\Delta=[\delta_1\geq \delta_2\geq\ldots\geq \delta_{l_\Delta}>0]= [\ \ldots,2^{m_2}, 1^{m_1}]$,
with $z_\Delta = \prod_k k^{m_k}\cdot m_k!\ $
It is well known, that Schur functions $\chi_R\{p\}$ form another orthonormal basis
when all $g_i=1$,
\be
\Big< \chi_R| \chi_{R'} \Big>^{(I)} =\delta_{R,R'}
\ee
and Kerov functions are defined as their triangular transformation,
which diagonalizes the product for arbitrary $\{g\}$:
\be
{\rm Ker}^{(g)}_R\{p\} = \chi_R\{p\} + \sum_{R'<R} { K}^{(g)}_{R|R'}\cdot\chi_{R'}\{p\}\nn \\
\Big< {\rm Ker}^{(g)}_R| {\rm Ker}^{(g)}_{R'} \Big>^{(g)} =
||{\rm Ker}_R^{(g)}||^2\cdot \delta_{R,R'}
\ee
It is triangular
w.r.t. lexicographical ordering of the Young diagrams of the same size,
for $R=[r_1\geq r_2\geq \ldots \geq r_{l_R}>0]$
\be
R>R'  \ \ {\rm if} \ \ r_1>r_1' \ \ {\rm  or\ if} \ \ r_1=r_1', \ {\rm but} \ r_2>r_2',
\ \ {\rm or\  if} \ \ r_1=r_1' \ {\rm  and} \ r_2=r_2', \ {\rm but} \ r_3>r_3',
\ \ {\rm and\ so\ on}
\label{lexico}
\ee
Alternative ordering, when lexicographically ordered are transposed Young diagrams,
gives risen to a non-equivalent set of dual Kerov functions, and both sets nicely complement
each other, for example in Cauchy formula and all its far-going implications.
Actually parameters $g_k$ appear in many formulas through Schur functions,
depending on them as on the new, additional set of time-variables,
what can imply an essentially new twist in the theory of integrable systems.
The best-known particular example of Kerov functions is provided by the set of
Macdonald polynomials, associated with the strangely-looking choice
\be
g^{\rm Mac}_n = \frac{\{q^n\}}{\{t^n\}}
\ee
Empirically this choice is distinguished by a tremendous simplification of many formulas,
especially those for the Littlewood-Richardson  coefficients,
what allows to preserve close relation to conventional representation theory.
However, an {\it a priori} reason for this simplification,
at the level of triangular structures, remains an open puzzle.

The scalar-product approach is technically very powerful,
however it has a conceptual drawback:
there is no explanation why original basis, which needs to be triangularly
transformed, consists of Schur functions
rather than any other set, obtained by some orthogonal transformation,
which always preserves both the scalar product and Cauchy formula.

\bigskip

{\bf 2.} The next example is provided by generalized Macdonald functions (GMF) \cite{GenMacrefs},
which play the crucial role  in description of AGT relations \cite{AGT}.
In the simplest two-times case they are triangular transforms from the factorized basis
$M_{R_1}\{p\}\cdot M_{R_2}\{\bar p\}$:
\be
{\cal M}_{R_1,R_2}\{p,\bar p\} = M_{R_1}\{p\}\cdot M_{R_2}\{\bar p\}
+ \sum_{(R_1,R_2')<(R_1,R_2)} {\cal C}_{R_1,R_2|R_1',R_2'}(A,q,t) \cdot
M_{R_1'}\{p\}\cdot M_{R_2'}\{\bar p\}
= \\
= \chi_{R_1}\{p\}\cdot \chi_{R_2}\{\bar p\}
+ \sum_{(R_1,R_2')<(R_1,R_2)} {C}_{R_1,R_2|R_1',R_2'}(A,q,t)\cdot
\chi_{R_1'}\{p\}\cdot \chi_{R_2'}\{\bar p\}
\ee
with triangular ${\cal C}$ and $C$ depending on a deformation parameter $Q=A^2$.
Ordering for the pairs of equal size, $|R_1|+|R_2| = |R_1'|+|R_2'|$, is defined by the rule
\be
(R_1,R_2)> (R_1',R_2') \ \ \ {\rm if} \ \ \ |R_2|>|R'_2| \ \ \ {\rm or} \ \ \
|R_2|=|R'_2| \ \ \ {\rm and} \ \ \ R_2>R_2' \ \ \ {\rm or} \ \ \
R_2=R'_2 \ \ \ {\rm and} \ \ \ R_1>R_1'
\label{doublexico}
\ee
GMF with  $n$ time sets are defined in a similar way,
and triangular is
the deformation of the factorized basis $\prod_{a=1}^n M_{R_a}\{p^{(a)}\}$.
\vspace{0.2cm}

GMF have no nice definition in terms of the scalar products
(the standard approach
exploits orthogonality
to {\it another} basis -- of the "dual" GMF),
see \cite{MMgenmac} for
discussion and references.
Instead they are specified in a different way --
as eigenfunctions of the triangularly-deformed Hamiltonians \cite{Hamsord},
the simplest one being
\be
\hat {\cal H} (F)\{p,\bar p\} = \frac{1}{t^2-1}\left\{-(1+Q^{-1})\cdot  F\{p,\bar p\} +
{\rm res}_{z=0}\left(
\exp\left(\sum_n \frac{(1-t^{-2n})p_nz^n}{n}\right)
F\left\{p_k+\frac{q^{2k}-1}{z^k},\bar p_k\right\}
+ \right.\right. \nn \\ \!\!\!\!\!\!\!\!\!\!\!\!\! \left.\left.
+ Q^{-1}\cdot  \exp\left(\sum_n \frac{(1-t^{-2n})z^n}{n}
\Big(\underline{\left(1-(t/q)^{2n}\right)p_n}+\bar p_n)\right)
F\left\{p_k,\bar p_k + \frac{q^{2k}-1}{z^k} \right\}
\right)\right\} \ \ \ \ \
\label{GEMham}
\ee
It is triangular in the sense that the mixing term (underlined)
between the ordinary  Ruijsenaars Hamiltonians
lowers the level $|R_2|$ of $M_{R_2}\{\bar p\}$
and trades this for increase of the level $|R_1|$ of $M_{R_1}\{p\}$.

These are well known, still somewhat mysterious Hamiltonians.
Hamiltonian approach has an advantage that it selects the basis,
in particular, distinguishes Schur functions $\chi\{p\}$
from monomials $p^\Delta$, but instead it is considerably
less efficient technically.
And even conceptually it is not always revealing triangular structures
in explicit way.

\bigskip

{\bf 3.}
In particular, we can refer to sec.1 and wonder why the Ruijsenaars Hamiltonian,
\be
\hat{\cal H} = \oint \frac{dz}{z}   \exp\left(\sum_n \frac{\{t^n\}\,p_nz^n}{n}\right)
\exp\left(\sum_n \left(\frac{q}{t}\right)^n\frac{\{q^n\}}{z^n}\frac{\p}{\p p_n}\right)
\ee
defines the { ordinary} Macdonald polynomials, which are its eigenfunctions,
as {\it triangular} transformations of Schurs.
Here and below we use the standard notation $\{x\} = x-x^{-1}$.
We give just an example of how it works, with the help of Cauchy formula,
\be
\exp\left(\sum_n \frac{p_np_n'}{n} \right) = \sum_Y \chi_Y\{p\}\chi_Y\{p'\}
\ee
recently reviewed for these purposes in \cite{Mcauchy}.
In our case the role of $p'$ will be played by a peculiar set $p'_k=\{t^k\}$,
and we use the fact that Schur functions at this locus are non-vanishing
only for the single-hook diagrams $Y = [a+1,1^b]$:
\be
\chi_{_{[a+1,1^b]}}\big\{p_k=\{t^k\}\big\} = (-)^b\cdot t^{a-b}\cdot \{t\}
\ee
Therefore
\be
\exp\left(\sum_n \frac{\{t^n\}  p_n z^n}{n}\right)
= \sum_{a,b} (-)^b \cdot z^{a+b+1}\cdot t^{a-b}\cdot \{t\}\cdot
\chi_{_{[a+1,1^b]}}\{p\}
\nn\\
\exp\left( \sum_n \left(\frac{q}{t}\right)^n \frac{\{q ^n\}}{z^n} \frac{\p}{\p p_n}\right)
\chi_{_R}\{p\}
= \sum_{a,b}(-)^b \cdot \left(\frac{q}{tz}\right)^{a+b+1}\cdot q^{a-b}
\cdot  \{q\}\cdot\chi_{_{R/[a+1,1^b]}}\{p\}
\ee
where $\chi_{_{R/Y}}$ are the {\it skew} Schur functions, arising when  an operator
$\chi_{_Y}\!\left\{k\frac{\p}{\p p_k}\right\}$ acts on $\chi_{_R}\{p\}$.
Thus
\be
\frac{\hat {\cal H} - 1}{\{t\}\{q\}}\, \chi_{_R} =
\frac{q}{t}\cdot \chi_{[1]}\chi_{_{R/[1]}}
+ \left(\frac{q}{t}\right)^2\left(t\cdot\chi_2-\frac{1}{t}\cdot\chi_{_{[1,1]}}\right)
\left(q\cdot\chi_{_{R/[2]}}-\frac{1}{q}\cdot\chi_{_{R/[1,1]}}\right) + \ldots
\ee
For  $R=[1,1]$ we get:
\be
\frac{q}{t}(\chi_{[2]}+\chi_{[1,1]}) -\frac{1}{q}
\left(\frac{q}{t}\right)^2\left(t\cdot\chi_{[2]}-\frac{1}{t}\cdot\chi_{[1,1]}\right) =
\frac{q\,(1+t^{-2})}{t}\cdot\chi_{[1,1]}
\ee
For $R=[2]$ the eigenfunction is a triangular transform $\chi_{[2]}+\alpha \chi_{[1,1]}$.
Then we get
\be
(1+\alpha)\frac{q}{t}(\chi_{[2]}+\chi_{[1,1]}) +\left(q-\frac{\alpha}{q}\right)
\left(\frac{q}{t}\right)^2\left(t\chi_{[2]}-\frac{1}{t}\chi_{[1,1]}\right) =
\frac{q }{t}\left((1+q^2)\chi_{[2]}
+ \left(1-\frac{q^2}{t^2}+\alpha+\frac{\alpha}{t^2}\right)\chi_{[1,1]}\right)
\ee
and this is an eigenfunction, provided
\be
1-\frac{q^2}{t^2}+\alpha+\frac{\alpha}{t^2} = \alpha\cdot  (1+q^2) \ \ \
\Longrightarrow \ \ \ \alpha = \frac{\{q/t\}}{\{qt\}}
\ee
what is exactly the coefficient, defining the Macdonald function
$M_{[2]}\{p\} \sim \chi_{[2]}\{p\} + \frac{\{q/t\}}{\{qt\}}\chi_{[1,1]}\{p\}$. \\
Note that in the case of Schur functions, this $\hat {\cal H}$ is actually
not a single operator, but an entire one-parametric
family of commuting Hamiltonians, depending on a free parameter $q=t$ --
and no higher Hamiltonians are actually needed.
However, in Macdonald and generalized Macdonald cases they get more important.
In the most interesting realization they involve
skewing over diagrams with a larger number of hooks.

The simple example in this section demonstrates that triangularity
is not always obvious at the Hamiltonian level.
Even more difficult is the inverse problem -- to construct the deformation of Hamiltonians
from the triangular deformation of eigenfunctions.
In particular the generalization of Hamiltonians from Macdonald to Kerov functions
and thus the very {\it definition} of generalized Kerov functions  remain open questions.
A possible new insight can be provided by a very different appearance of
triangular Hamiltonian, which comes supplemented by additional "pentad" structure,
which, if better understood, can serve as a tool to fixing numerous ambiguities
in the triangular-transformation (dressing) approach.

\bigskip

{\bf 4.}
Pentad structure was discovered in the study of double-twist knots
\cite{doublebraidfirst}--\!\!\cite{doublebraidlast},
we refer to \cite{doublebraidlast} for a summary, pictures and notation.
These are the simplest arborescent knots,
with {\it reduced} colored HOMFLY-PT polynomials \cite{knotpols}
given by a very simple formula \cite{arbor}
\be
H_R^{(m,n)} = d_R \cdot
\Big<\emptyset\Big|\bar S \bar T^{2m} \bar S \bar T^{2n} \bar S\Big|\emptyset\Big>
\label{arbor}
\ee
where $\bar T$ is the colored ${\cal R}$-matrix in representation $R$,
while $\bar S$ is the exclusive Racah matrix ($6j$-symbol),
relating the two maps of representation products
$(R\otimes \bar R) \otimes R  \longrightarrow R$ and
$R\otimes (\bar R \otimes R)  \longrightarrow R$.
Dependence of knot invariants on "evolution parameters" $n$ and $m$
is controlled by the eigenvalues of $\bar T$,
and the conventional technique \cite{evo} is to work in the basis,
where $\bar T$ is diagonal.
Then the only non-trivial ingredient is the symmetric and orthogonal
Racah matrix $\bar S=\bar S^{\rm tr}$, $\bar S^2=I$.
It heavily depends on $R$, but we suppress the label $R$ in $\bar S_R=\bar S$
to make the formulas readable.
The problem is that it is not actually known from representation theory
and somewhat difficult to calculate from the first principles.
Thus what happened is that instead of being used to calculate
HOMFLY-PT polynomials, $\bar S$ was instead {\it deduced} from the
intuition about the twist knots.
The benefit for know theory is that once known, the same $\bar S$
can be used to calculate colored HOMFLY-PT polynomials for all other arborescent knots
and links.
Moreover, the insights about Racah matrices, revealed in this way,
can probably be used for more general mixing matrices \cite{AMMmix},
where they can be also combined with the powerful eigenvalue hypothesis
\cite{evhyp}.

The main knot-theory implication is the differential expansion \cite{diffexpan},
implying the separation of representation $R$ and knot/link ${\cal K}$ dependencies
\be
H^{\cal K}_R =  \sum_X Z_R^X F_X^{\cal K}
\label{diffexpan}
\ee
where $X$ are Young diagrams, somehow restricted (through vanishing of the $Z$-factors)
for the given $R$.
It appears that the interplay between (\ref{arbor}) and (\ref{diffexpan})
for the double twist family is highly non-trivial --
and  is actually enough to obtain explicit expression for Racah matrix $\bar S$.
The procedure is fully described in \cite{doublebraidlast} for all rectangular $R$,
where there are no multiplicities in the product $R\otimes\bar R$,
but it actually remains the same for arbitrary $R$, some technical details
still remain to be worked out in this case.
The main discovery  \cite{KNTZ} is that to rewrite (\ref{arbor})
in the form (\ref{diffexpan})
one needs to actually refrain from diagonalizing $\bar T$:
instead of diagonal "Hamiltonian" one needs to use {\it triangular} --
and this is what makes the story resembling that in s.2 above.
However, this time we can see additional miracles and more structures.

To be more precise, we substitute evolution with the diagonal matrix $\bar T$
by that with the triangular KNTZ matrix ${\cal B}$  \cite{KNTZ,doublebraidlast}.
At the first step we rewrite (\ref{arbor}) by insertion of unity decompositions
with the help of an auxiliary matrix $U$:
\be
\Big<\emptyset\Big|\bar S \bar T^{2m} \bar S \bar T^{2n} \bar S\Big|\emptyset\Big>
= \Big<\emptyset\Big|
\overbrace{ U^{\rm tr}
\,\Big(\!(U^{\rm tr})^{-1}}^I \bar S \bar T^{2m+2} \overbrace{\bar S U^{\rm tr}\Big)
\Big(\!(U^{\rm tr})^{-1}\bar S}^{I} \bar T^{-2} \bar S \bar T^{-2} \overbrace{\bar S U^{-1}\Big)
\Big(U\bar S}^I \bar T^{2n+2} \bar S \overbrace{U^{-1}\Big)U}^I\Big|\emptyset\Big>
\ee
Then the fact is that $U$ can be {\it adjusted} in such a way that the five facts are
simultaneously true:
\begin{itemize}

\item{
${\cal B} := U \bar S \bar T^2 \bar S U^{-1}
$ is triangular:
${\cal B}_{XY} \neq 0 \ {\rm iff}\ Y\subset X$
}

\item{
$U_{X\emptyset} = 1 \ {\rm or}\ \, 0$
}

\item{
$\Big<X\Big| {\cal B}U \Big|\emptyset\Big>=\delta_{X,\emptyset}$
}

\item{
$Z$-factor in (\ref{diffexpan}) is  given by a matrix element
$Z_R^X := d_R\cdot
\Big<\emptyset\Big| \bar S  \bar T^{2} \bar S \bar T^{-2}\bar S U^{-1}\Big|X\Big>
$
}

\item{
$d_R\cdot \Big<X\Big|(U^{\rm tr})^{-1}\bar S  \bar T^{-2} \bar S \bar T^{-2}
\bar S U^{-1}\Big|Y\Big>
=\frac{Z_R^X}{\Lambda'_X}\cdot \delta_{X,Y}
$
is diagonal matrix, with the same $Z^X_R$
}

\end{itemize}

Taken together, these facts mean that (\ref{arbor}) is indeed rewritten in the form
(\ref{diffexpan}) with the original mysterious factorization conjecture \cite{doublebraidfirst}
\vspace{-0.4cm}
\be
F^{(m,n)}_X = \frac{F^{(m)}_XF^{(n)}_X}{\Lambda'_X}
\ee
with $\Lambda'_X = F_X^{(1)} = {\cal B}_{X\emptyset}$ -- a generically known monomial
$\Lambda'_X = q^{\alpha_X}\cdot A^{\beta_X}$,
and the triangular evolution formula \cite{KNTZ} for the twist-family coefficient
\be
F_X^{(m)} = \Big<X\Big| {\cal B}^{m+1} U\Big|\emptyset\Big> =
\sum_Y \Big<X\Big| {\cal B}^{m+1} \Big|Y\Big>
\ee

{\bf 5.} Explicit example of the double twist family actually allows to understand
what is the set $X$ in (\ref{diffexpan}) -- then by universality of
the differential expansion this remains true for arbitrary knots.
Thus $X$ are actually the elements of the representations product $R\otimes \bar R$
(remarkably, they are in one-to-one correspondence with those  of $R\otimes R$,
what allows the whole story about the differential expansion to be self-consistent).
In the case of $R\otimes\bar R$ these are actually {\it composite} diagrams
of the type $X=(\lambda,\lambda')$ with $\lambda,\lambda'\subset R$ and $|\lambda|=|\lambda'|$.
Moreover, for rectangular $R=[r^s]$ only diagonal composites with $\lambda'=\lambda$
contribute, what allows to label $X$ in (\ref{diffexpan}) in this case by
subdiagrams $\lambda$ of $R$.
In non-rectangular case there are also non-diagonal composites,
moreover, some diagrams appear with multiplicities -- and these two facts appear to be
intimately related, especially in the formulas for $\bar S$ and ${\cal B}$..
Still the entire construction remains the same.

\bigskip

{\bf 6.} The next crucial fact about it is that the matrix ${\cal B}$
is explicitly known, at least for arbitrary rectangular representations $R=[r^s]$,
where diagrams $X$ can actually be identified with the sub-diagrams of $R$:
\be
{\cal B}_{\lambda\mu} := (-)^{|\lambda|-|\mu|}\Lambda_\lambda\cdot
\frac{\chi_{_{\lambda^{\rm tr}/\mu^{\rm tr}}}^0
\chi_{_\mu}^0}{\chi_{_\lambda}^0} =
\boxed{
\Lambda_\lambda \cdot \frac{\chi_{_{\lambda/\mu}}\{-p^0_k\}\cdot\chi_{_\mu}\{p^0_k\}}
{\chi_{_\lambda }\{p^0_k\}}
}
\label{Brect}
\ee
It is expressed through $\Lambda:=\bar T^2$ and through the
skew Schur functions and thus is automatically  triangular.
Label "0" denotes restriction of  Schur functions to
\be
p_k = p_k^0 := \frac{\{q\}^k}{\{t^k\}}
\ee
with $t=q$.
In the second (boxed) version and with Schur substituted by Macdonald functions
the same formula provides a positive rectangularly-colored super-polynomial
\cite{KNTZ,doublebraidlast}.

For non-rectangular representations the story is a little more involved:
$\bar S$ splits into independent blocks,
with only one block contributing to arborescent  calculus.
The same is true for ${\cal B}$, which also acquires explicit $N$-dependence ($A=q^N$) in
additional rows and columns, and thus is not fully described by (\ref{Brect}).
For example, for $R=[2,1]$ the triangular matrix ${\cal B}$ is

\bigskip

{\footnotesize
\centerline{
$
\left(\begin{array}{c|cc||cc||cc|cc|cc|cc||cc}
1 &&&&&&&&&&&& \\
&&&&&&&&&&&& \\
-A^2&A^2 &&&&&&&&&&& \\
&&&&&&&&&&&& \\
\hline
&&&&&&&&&&&& \\
0&0&&A^2& &&&&&&&&&\\
&&&&&&&&&&&& \\
\hline
&&&&&&&&&&&& \\
-A^2&0&&0&&A^2 &&&&&&&\\
&&&&&&&&&&&& \\
\frac{A^4}{q^2} & -\frac{A^4}{[2]q^3}
&& \frac{A^4}{q^3[N]}\sqrt{\frac{[N+2]}{[N-2]}}
&& - \frac{[3]A^4}{[2]q^3} && \frac{A^4}{q^4} &&&&&&\\
&&&&&&&&&&&& \\
q^2A^4 & - \frac{q^3A^4}{[2]}
&& -\frac{q^3A^4}{ [N]}\sqrt{\frac{[N-2]}{[N+2]}}
&& -\frac{q^3[3]A^4}{[2]}&& 0 && q^4A^4 &&&&\\
&&&&&&&&&&&& \\
-A^6 & \frac{[3]A^6}{[2]^2}
&& -\frac{[3]A^5}{[N]\sqrt{[N+2][N-2]}}
&& \frac{[3]^2A^6}{[2]^2} && -\frac{[3]A^6}{q[2]}&& -\frac{[3]qA^6}{[2]}&&A^6 && \\
&&&&&&&&&&&& \\
-A^6&\frac{[3]A^6}{[2]^2}&\!\!\!\!\!\!\!\!-\frac{A^5[N]}{[2]\{q^2\}}
& -\frac{[3]A^5}{[N]\sqrt{[N+2][N-2]}} &\!\!\!\!\!+ \frac{A^6}{\{q\}\sqrt{[N+2][N-2]}}
&\frac{[3]^2A^6}{[2]^2}&\!\!\!\!\!\!\!\!\! +\frac{A^5[N]}{[2]\{q^2\}}
&-\frac{[3]A^6}{q[2]}&\!\!\!\!\!\!  +\frac{q^2A^5[N]}{ [2]}
&-\frac{[3]qA^6}{[2]}&\!\!\!\!\!\!  -\frac{A^5[N]}{q^2[2]}
&0&A^4 &\\
&&&&&&&&&&&& \\
\hline
&&&&&&&&&&&& \\
0&0&&0&&0&&0&&0&&0&0&A^4&0\\
0&0&&0&&0&&0&&0&&0&0&0&A^2
\end{array}\right)
$
}
}

\bigskip

\noindent
The $2\times 2$ block in the right low corner decouples,
and also vanishing is the
matrix element $U_{X\emptyset}$ with $X$ from the third line,
thus vanishing are sums of all the matrix elements along each line,
with omission of the ones in the third column.
All contributing $Z$-factors are nicely factorized, just as they are
for rectangular case, the non-factorized expression  in \cite{Ano21},
\be
Z_{[2,1]}^{[1]} \ = \  D_3D_{-3}+D_2D_0+D_0D_{-2} \ = \
\frac{[3]}{[2]^2}D_0^2 + \frac{[3]^2}{[2]^2} D_2D_{-2}
\ee
is actually a sum of two factorized $Z$-factors
from the second and the forth lines.

\bigskip

{\bf 7.}
As we saw, the matrix ${\cal B}$ is very simple.
Moreover,  it is {\it universal}, it does not actually depend on $R$,
specification of $R$ just cuts out a piece, restricted to relevant lines/columns $X$,
for which the $R$-dependent $Z$-factors are non-vanishing.

Matrices   $\bar S$ and $U$ are instead heavily $R$-dependent and rather complicated.
However, they are easily restored, if one knows another {\it universal triangular} matrix
${\cal E}$, made from the eigenfunctions of ${\cal B}$:
\be
{\cal B}{\cal E} = {\cal E}\Lambda \ \ \ \Longleftrightarrow \ \ \
{\cal B} = {\cal E}\Lambda{\cal E}^{-1}
\ee
Then ${\cal E} = U\bar S$ and
\be
d_R\cdot \bar T^{-2}{\bar S}\bar T^{-2} = \ {\cal E}^{\rm tr} \frac{Z}{\Lambda'} {\cal E}
\label{bSviaE}
\ee
Thus ${\cal E}$ diagonalizes the {\it triangular}  evolution operator ("Hamiltonian")
${\cal B}$, and at the same time it diagonalizes the Racah matrix ${\cal S}$,
but this time -- not as an operator, but as a quadratic form.
Instead, $\bar S$ can be diagonalized as an operator, and then the diagonalizing
matrix is the second exclusive Racah matrix $S$ \cite{arbor}:
\be
\bar T \bar S \bar T   = ST^{-1}S^{-1}
\ee
Thus we get a peculiar ordered pentad of matrices:
\be
&\boxed{
\begin{array}{ccccc}
\bar T &\\ \\
\downarrow & \\ \\
{\cal B} & \longleftrightarrow & {\cal E} \\
&&\\
&&\downarrow & \\
&&\\
&&\bar S & \longleftrightarrow & S
\end{array}
}
\ee
with one diagonal matrix in the first line,
two universal triangular matrices in the second line
and two $R$-dependent (non-universal) and non-triangular in the third line.
One can also add $U$ and $T$ to make a septet.

\bigskip

{\bf 8.} If ${\cal E}$ in the first line was just an eigenfunction matrix for ${\cal B}$,
it would be  ambiguous: one could multiply it by any diagonal matrix from the right.
Moreover, when $\bar T^2$ has coincident entries, the corresponding block can even be non-diagonal.
Such modification, however, affects orthogonality of $\bar S$, the normalization condition
$U_{X\emptyset}=1$ and the expressions for $Z$.
Any of them can be used to fix the ambiguity -- and this is why the pentad structure
seems to be more rigid than just the triangular one.
In fact, one ca rewrite the knot polynomial in a from which is free from the ambiguity,
\be
F_X^{(m)} = \sum_{Y:\ U_{Y\emptyset}=1} \Big<X\Big|{\cal E}\bar T^{2n} {\cal E}^{-1} \Big|Y \Big>
\ee
but there is no such non-ambiguous formula for $\bar S$.
Also, ambiguity in ${\cal E}$ could be used to get rid of the diagonal matrix
$Z/\Lambda'$ at the r.h.s. of (\ref{bSviaE}),  but this is also not so simple
the factor could be easily absorbed if the freedom was to multiply ${\cal E}$
from the left -- but instead we can do this only from the right.
Thus reconstruction of $\bar S$ from the known ${\cal B}$ is in fact a more delicate
procedure than that of building the HOMFLY-PT polynomial.
We illustrate it by the very simplest example of the fundamental representation $R=[1]$.
In this case:
\be
{\cal B} = \left(\begin{array}{cc} 1 & 0 \\ -A^2 & A^2 \end{array} \right), \ \ \ \
\bar T^2 = \left(\begin{array}{cc} 1 & 0 \\ 0 & A^2 \end{array} \right) \ \ \ \Longrightarrow \ \ \
{\cal E} = \left(\begin{array}{cc} 1 & 0 \\ \frac{A^2}{A^2-1} & c \end{array} \right)
\ee
where $c$ is the ambiguous parameter.
Also yet-unknown in this approach is   $Z'=Z_{[1]}/\Lambda'_{[1]}$ in (\ref{bSviaE}):
\be
\bar S = d_{[1]}\cdot\bar T^{2} {\cal E}^{\rm tr} \frac{Z}{\Lambda'} {\cal E} \bar T^{2} =
d_{[1]}\cdot\left(\begin{array}{cc} 1 & 0 \\ 0 & A^2 \end{array} \right)
\left(\begin{array}{cc} 1 &   \frac{A^2}{A^2-1}\\ 0  & c \end{array} \right)
\left(\begin{array}{cc} 1 & 0 \\ 0 & Z' \end{array} \right)
\left(\begin{array}{cc} 1 & 0 \\ \frac{A^2}{A^2-1} & c \end{array} \right)
\left(\begin{array}{cc} 1 & 0 \\ 0 & A^2 \end{array} \right)
= \nn
\ee
\vspace{-0.5cm}
\be
= d_{[1]}\cdot\left(\begin{array}{cc} 1+\frac{A^4Z'}{(A^2-1)^2} & \frac{cA^4Z}{A^2-1} \\
\frac{cA^4Z}{A^2-1} & c^2A^4Z \end{array} \right)
\ee
If we require this matrix to be orthogonal, we fix $c^{-1}= d_{[1]}\cdot A\sqrt{\{Aq\}\{A/q\}}$
and $Z'= -\frac{\{Aq\}\{A/q\}}{A^2}$.

Deriving a general formula for ${\cal E}$ is an open and challenging problem.
Even in the simplest case of the single-line (symmetric) $R$
explicit solution
\be
{\cal E}_{nm} = \frac{[n]!}{[m]![n-m]!}\prod_{j=n+m}^{2m-1}\{Aq^j\}\cdot c_m
\ee
satisfies ${\cal B}{\cal E}={\cal E}\bar T^2$ due to
rather exotic combinatorial identity:
\be
q^{2Nn}\cdot    \sum_{i=0}^n \big(-\{q\}\big)^{ i} \cdot
\frac{[n]!}{[n-i]![i]!}\cdot \frac{[N !}{[N-i]!} \cdot q^{\frac{i(i+1)}{2}-(N+n)i} = 1
\ee
which is a $q$-deformation of
\be
 A^n \cdot \sum_{i=0}^n (-)^{n-i}\frac{n!}{(n-i)!\,i!} \,(A-A^{-1})^{n-i} A^i  =
A^n \cdot\Big(A-(A-A^{-1})\Big)^n = 1
\ee
Since Racah matrix $\bar S$ is quadratic in ${\cal E}$,
explicit knowledge of this matrix is important to generalization of hypergeometric
series \cite{hypergeom} for $\bar S$ and $S$ from symmetric to arbitrary representations $R$.

\bigskip

{\bf 9.} If we treat like ${\cal B}$ the Hamiltonians ${\cal H}$ from sec.2 and 3,
then ${\cal E}$ is the matrix of their
eigenfunction, i.e. for Macdonald-Kerov theory from secs.1 and 2
these would be Macdonald, Kerov or generalized Macdonald functions
(more precisely, ${\cal E}$ would be inverse of triangular Kostka matrices $K$).
The question is what are the other three matrices $\bar S$, $S$ and $U$.
The first of them, $\bar S$ looks like the matrix of scalar products for Schurs,
provided Macdonald/Kerov functions are orthogonal.
The rigidity of pentad structure can help to explain what distinguishes
particular choices of basises for symmetric functions and especially for GMF.
Hopefully this can also help to resolve the long-standing problems
of Kerov Hamiltonians and higher Hamiltonians for GMF,
and all the way further, to generalized Kerov functions.


\section*{Acknowledgements}

I am indebted for discussions to A.Mironov, A.Sleptsov and Y.Zenkevich.
This work was partly supported by the Russian Science Foundation (Grant No.16-12-10344).

\end{document}